\documentclass[aps,pre,showpacs,twocolumn]{revtex4}
\usepackage{amssymb}
\usepackage{graphicx}
\usepackage{colordvi}
\usepackage{color}
\usepackage{dcolumn,amsmath,amsthm,amscd,amsfonts,amssymb,epsfig,graphics,graphicx,eucal}

\newcommand{\be}{\begin{equation}}
\newcommand{\ee}{\end{equation}}

\begin{document}

\title{Scattering properties of a ${\cal{PT}}$  dipole}
\author{K. Staliunas$^{{1},{2}}$,   P. Marko\v s$^{3}$,  and  V. Kuzmiak$^{4}$}
\affiliation{$^1$Department de F{\'i}sica i Egineria Nuclear, Universitat Polit{\` e}cnica de Catalunya (UPC), Barcelona, Spain
\\
$^2$Instituci{\' o} Catalana de Recercai Estudis Avançats (ICREA), Barcelona, Spain
\\
$^3$Department of Experimental Physics, Faculty of  Mathematics Physics and Informatics, Comenius University in Bratislava, 842 28 Slovakia
\\
$^4$  Institute of Photonics and Electronics, Academy of Sciences of the Czech Republic,v.v.i., Chaberska 57, 182 51, Praha 8, Czech Republic
}

\date{\today}

\begin{abstract}
Electromagnetic response of
${\cal{PT}}$-dipole
is studied both analytically and numerically. In analytical approach, dipole is represented by two point scatterers. Within the first Born approximation, the asymmetry of the scattering field with respect to the orientation of dipole is proven. In numerical simulations, dipole is represented by two infinitely long, parallel cylinders with opposite sign of imaginary part of refractive index. Numerical data confirm the validity of the Born approximation in the weak scattering limit, while significant deviations from the Born approximation were observed for stronger scatterers and in the near-field range.
\end{abstract}
\maketitle

\input{epsf.tex}
\epsfverbosetrue

\section{Introduction}
\label{intro}



Parity-time ($\cal{PT}$) symmetry has recently emerged as a promising design principle for extending Hermitian to non-Hermitian optics and has given rise to a rich variety of physical phenomena based on the appearance of exceptional points and phase transitions in the eigenvalues of the associated non-Hermitian Hamiltonian \cite{bender1,bender2}.
In classical wave systems, where real part of the potential in optics is the refractive index and gain/loss is analogous to its imaginary part and $\cal{PT}$-symmetry demands that $n(r) = n^{*}(-r)$, one can envisage various structures obtained by combining the index and gain/loss modulations with required symmetries which represent classical analogues of quantum systems described by $\cal{PT}$-symmetric potentials.
 The recent experimental realizations of $\cal{PT}$-symmetric optical systems have attracted widespread interest, in particular due to their promising prospect to achieve tunable components with extreme sensitivity and very unconventional wave behavior\cite{iop}-\cite{guo}. These include loss induced invisibility\cite{lin},  Bloch oscillations\cite{longhi1}, laser generation by reversing the effect of loss at threshold \cite{peng}-\cite{phang1}, unidirectional propagation\cite{regen}-\cite{feng1}, optical solitons in PT periodic systems \cite{mussli}-\cite{wimmer}, to name a few of numerous new concepts proposed.

The key feature of $\cal{PT}$ symmetric photonic structures stems from the fact that they may have real eigenvalues despite having gain and loss which break the space symmetry. For a certain amount of gain/loss, there exists a threshold at which the system undergoes a spontaneous $\cal{PT}$-symmetry breaking, and above which eigenfrequencies become complex and power grows exponentially. Based on the $\cal{PT}$ concept several types of extended systems included $\cal{PT}$ gratings, $\cal{PT}$ lattices and $\cal{PT}$-symmetric resonant structures characterized by the complex-valued periodic functions have been studied both theoretically and experimentally\cite{feng2}-\cite{phang2}. The periodic systems also provide asymmetric response, offering for example an unidirectional invisibility \cite{lin}, unidirectional coupling \cite{kestas1} and other peculiar effects. Due to periodicity such systems are resonant, i.e. provide the asymmetric responses in the vicinity  resonant wavelength $\lambda \sim 2a$, where and $a$ is the period of the structure.

Most of the $\cal{PT}$ studies focus on one-dimensional systems. A naive extension to 2D is possible by considering parity symmetry in one space direction, which however can hardly lead to principally novel effects. Some exception is perhaps in \cite{kestas1}, where the nontrivial chiral-$\cal{PT}$ concept has been introduced, which is possible only in 2D or 3D systems. Majority of the studies also consider the global $\cal{PT}$ -symmetry. This means that the bulk is uniformly filled by a continuous $\cal{PT}$-media (i.e. by $\cal{PT}$-lattices). Perhaps a single exception within this context is the work \cite{kestas2}, where the local-$\cal{PT}$ concept has been introduced for the first time, providing the $\cal{PT}$ effect in different directions. This results in $\cal{PT}$-systems with nontrivial flows, i.e. for instance axisymmetric flows toward some focus as suggested in original proposal \cite{kestas2}. Moreover, it could be extended to build a systems with arbitrary $\cal{PT}$-flows, like closed loops of currents, chiral objects, and any other flow-configuration on demand \cite{kestas3}.

\begin{figure}[t]
\noindent\includegraphics[width=0.3\textwidth]{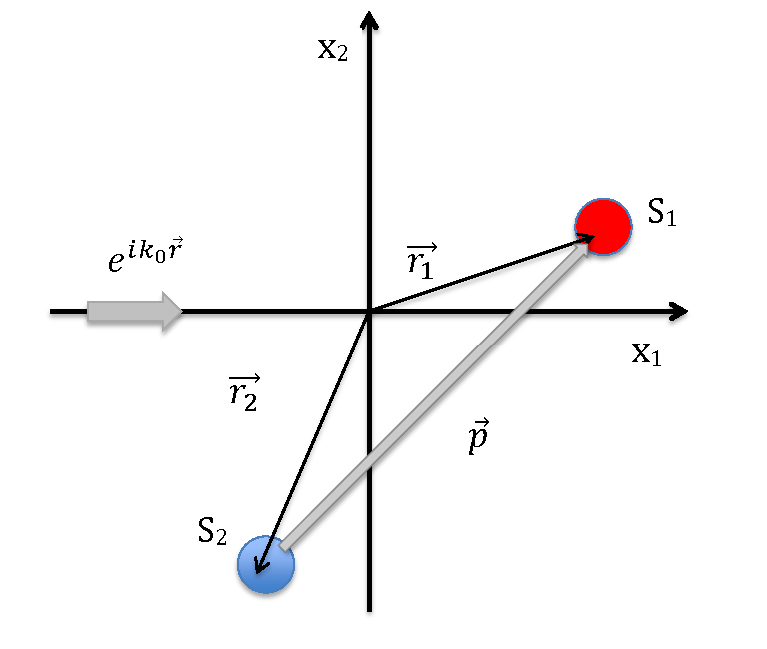}
\caption{(Color online) A single ${\cal{PT}}$-dipole.}
\label{fig_dipole}
\end{figure}

The most natural way of looking into the physical properties of such complicated $\cal{PT}$-objects is to consider them as structures built from microscopic $\cal{PT}$ objects i.e. so called $\cal{PT}$ molecules represented by $\cal{PT}$-dipoles – which can be described as generalized form of {\color{black}the} conventional ones  - see Fig. \ref{fig_dipole}. The main idea behind our paper is to identify $\cal{PT}$-dipole as a minimum unit building block which possesses the $\cal{PT}$ properties. We demonstrate that such a minimum object consists of two scatterers with different complex scattering coefficients.
We demonstrate that such a minimum object consists of two scatterers with different complex coefficients. Realization of PT-symmetry in optics requires considerable amounts of gain loss can be provided only by semiconductors and polymers. For example, $\cal{PT}$-symmetry breaking was observed experimentally in a passive $\cal{PT}$-symmetry ridge optical waveguide consisting of multilayer Al$_x$Ga$_{1-x}As$ heterostructure with varying concentration, where the loss is introduced though deposition of a thin layer of chromium on one of the coupler arms\cite{guo}. Another configuration employing the $\cal{PT}$ symmetry concept was demonstrated in the $\cal{PT}$-synthetic microring resonator with InGaAsP multiple quantum wells deposited on InP substrate where balanced gain/loss modulation is achieved by periodically formed Cr/Ge structures on the top of the InGaAsP\cite{feng2}.  A possible design of the $\cal{PT}$-dipole that could be implemented and measured in microphotonic devices was proposed in the context of 2D $\cal{PT}$-symmetric complex structure\cite{kestas1}. The configuration shown in Fig. 4(a) in Ref.~\onlinecite{kestas1} consists of a dielectric slab with holes filled by p/n and n/p semiconductor junctions which provide gain or loss depending on the orientation {\color{black}of} each component.

Our paper is organized as follows. In Sec. II we develop a scattering theory of a $\cal{PT}$-dipole within the first Born approximation and define the difference in intensity of scattered field between the configurations with $\cal{PT}$ dipoles parallel($\vec{p}$) and antiparallel ($-\vec{p}$) to the direction of the incident wave.  We consider two specific configurations of the $\cal{PT}$ dipole aligned parallel and perpendicular with respect to the incident wave. In Sec. III we numerically investigate the scattering properties of the $\cal{PT}$ -dipole represented by the system consisting of infinitely long, parallel cylinders with the opposite sign of the imaginary component of the refractive index. In Sec. IV we present the numerical results for both configurations of the $\cal{PT}$-dipole considered. The discussion of the validity of theoretical model based on the first Born approximation and deviations identified by numerical approach are discussed in Sec. V.

\section{Theoretical model}
\label{theory}

\def\sr{S_{\textrm{\footnotesize{Re}}}}
\def\si{S_{\textrm{\footnotesize{Im}}}}

\subsection{The model}

The ${\cal{PT}}$ dipole consists of two point scatterers centered at the positions $\vec{r}_1$ and $\vec{r}_2$  in the $xy$ plane with different complex scattering  coefficients $s_1$ and $s_2$ -- see Fig.~\ref{fig_dipole}. The distance between two scatterers, $a = |\vec{r}_1-\vec{r}_2|$ defines the length scale of the model.
Scattering coefficients which represent effective polarizabilities can be in general complex,
\be s_{1,2} = \sr\pm i~\si.
\ee
The real component corresponds to elastic scattering and the  imaginary one accounts for the emission/absorption.
The incident field is the  plane wave propagating in the $xy$ plane with unit amplitude and normalized frequency $f = a/\lambda$.

\subsection{Electromagnetic response of
the ${\cal{PT}}$ dipole}

The total electric field $E$ at a point $\vec{r}$ is assumed to be parallel to the $z$ axis can be written
as a superposition of incident plane wave with unit amplitude and  field $E_S$ scattered by two scatterers:
\be
E(\vec{r}) = e^{i\vec{k}_0\vec{r}} +  E_S(\vec{r}).
\ee
Here,  $k_0= 2\pi/\lambda$ is the wave vector of the incident wave.

In the limit of weak scattering one can use the first Born approximation which allows to calculate the field far away from the scattering center. To describe the behavior of a single dipole consisting of two point scatterers we focus on the scattered field that can be written as
\begin{equation}
E_{\rm S}(\vec{r}) = \sum_{j=1}^{2}
\frac{  i s_j e^{i\vec{k}_0\vec{r}_j}e^{i |\vec{k}_0||\vec{r}-\vec{r}_j|}}
{|\vec{r} - \vec{r}_j|^{1/2}}.
\label{E_scatt_Born}
\end{equation}
In order to obtain analytical insight we first simplify the general expression for the scattered field  into asymptotic form in the far-field limit assuming  $|\vec{r}| \gg |\vec{r}_{1,2}|$
\begin{equation}
E_{\rm S}(\vec{r}) = \frac{ ie^{i |\vec{k}_0||\vec{r}|}}
{ |\vec{r}|^{1/2} }
\left(\sum_{j=1}^{2} s_j e^{ i |\vec{k}_0|(\vec{e_k}-\vec{e_r})\vec{r_j}}  + O(|\vec{r}|^{-1})\right)
\label{E_scatt_far-field}
\end{equation}
where unit vectors $\vec{e_k} = \vec{k}_0/|\vec{k}_0|$ and $\vec{e_r} = \vec{r}/|\vec{r}|$ indicate directions of the incident wave and of the observer at the point $\vec{r}$, respectively.

\medskip

To understand how  the ${\cal{PT}}$ symmetry influences the electromagnetic response of the dipole, we first consider the  limit of ``small'' dipole $k_0a \ll 1$ the latter equation can be simplified into the form
\begin{equation}
E_{\rm S}(\vec{r}) = \frac{ ie^{i |\vec{k}_0||\vec{r}|}}
{ |\vec{r}|^{1/2} }
\left( s + i (\vec{e_k}-\vec{e_r})\vec{p}  + O(|\vec{r}|^{-1})\right)
\label{E_scatt_small-dipole}
\end{equation}
The first term in the bracket in the latter equation represents the total scattering defined as a sum of the scattering coefficients associated with each of the scatterers
$s = \sum_{j} s_j$. This term is parity-invariant and thus provides symmetric scattering, while the second term in which
\be\vec{p} = |\vec{k}_0|\sum_{j}{\color{black}s_j}\cdot\vec{r_j}\ee
defines the $\cal{PT}$ dipole  gives rise to asymmetry of scattering which depends on both its strength and orientation.

The parity asymmetry of the scattering can be determined by calculating the scattered field for the dipoles with opposite orientations $\vec{p}$ and $-\vec{p}$.
Specifically, in the forward direction $\vec{e_k} = \vec{e_r}$ when orientation of the dipole coincides with the direction of the incident wave, the second term in the brackets of the Eq.(\ref{E_scatt_small-dipole}) vanishes and the scattered field does not depend on the $\cal{PT}$ dipole $\vec{p}$. In the case of backward scattering when $\vec{e_k} = -\vec{e_r}$ the scattered field given by Eq. (\ref{E_scatt_small-dipole}) is proportional to $s + 2i\vec{e_k}\vec{p}$. This means that for real-valued scalar scattering $s = \sum_{j} s_j$ and for real-valued $\cal{PT}$ dipole $\vec{p}$ {\color{black}the} parity symmetry is maintained i.e. $|s + 2i\vec{e_k}\vec{p}| = |s - 2i\vec{e_k}\vec{p}|$. On the other hand, when the scattering is described in  terms of the complex coefficients
${\color{black}s_j,}$ the parity symmetry is broken.
Thus, asymmetric scattering requires the dipole characterized by nonzero elastic scattering and a nonzero gain/loss balance.

\medskip

\begin{figure}[t]
\noindent\includegraphics[width=0.4\textwidth]{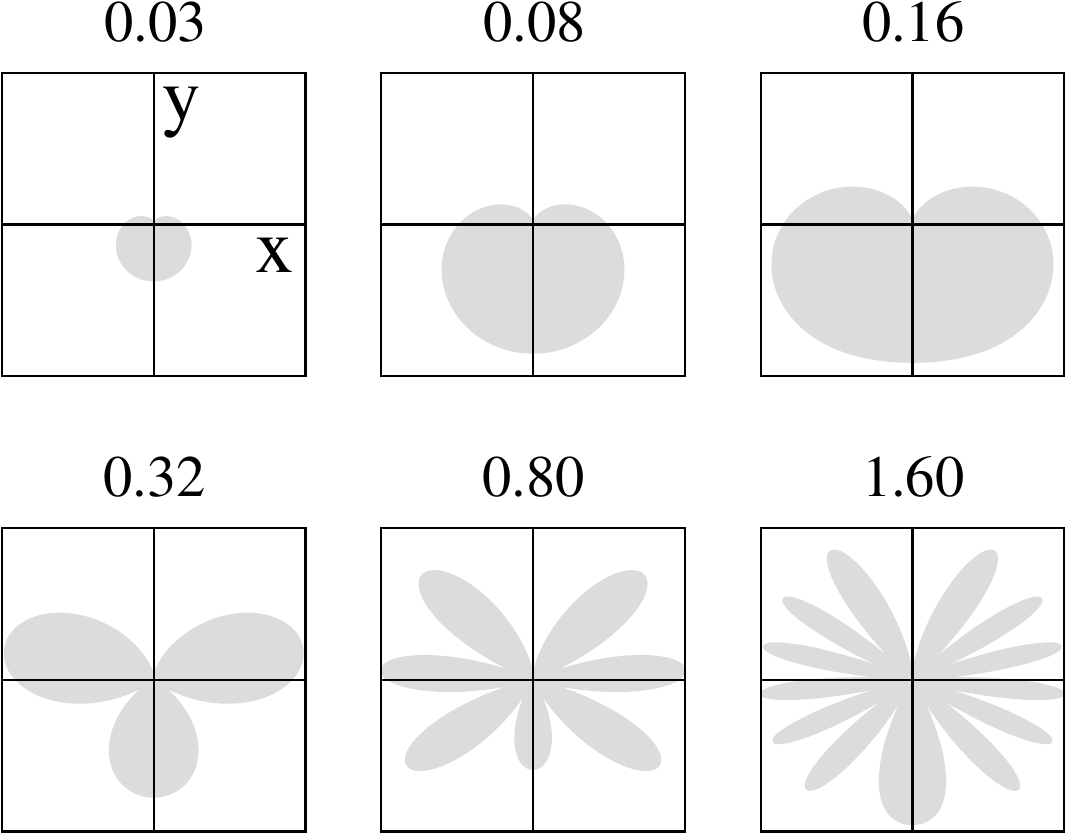}    
\caption{Angular diagrams of $\cal{PT}$ asymmetry $\Delta P_{S}(\vec{r})$ in the case of parallel orientation of the $\cal{PT}$-dipole for the normalized frequencies $f=a/\lambda = k_0a/(2 \pi)$ in the range $0.03 < f < 1.6$. Incident wave propagates along the $y$ axis. Orientation of axes is shown in the left upper panel. Since  all panels have the same scale, diagrams give also an estimation how the total scattered energy depends on the frequency of incident wave.}
\label{fig_1}
\end{figure}

The preceding discussion is valid also without assumption of ``small'' dipoles.  In what follows the condition $k_0a\ll 1$ is lifted.
By using the notation $\vec{r}_{1,2} = \pm \Delta \vec{r}/2$, the Eq.(\ref{E_scatt_far-field}) can be rewritten into the form
\begin{equation}
\label{E_scatt_dipole}
\begin{array}{lcl}
\displaystyle{E_{\rm S}(\vec{r},\vec{p})  = \frac{ 2ie^{i |k_0||\vec{r}|}}
{ |\vec{r}|^{{1}/{2}} } }
\Big[ &&\sr\cos(\frac{|\vec{k}_0|(\vec{e_k}-\vec{e_r})\cdot\Delta\vec{r}}{2}) \\
&+&
\si\sin(\frac{|\vec{k}_0|(\vec{e}_k-\vec{e}_r)\cdot\Delta\vec{r}}{2})\Big],
\end{array}
\end{equation}
according to which the $E_S(\vec{r},\vec{p})$ depends on the orientation of the dipole.
To characterize the asymmetry of scattered field associated with the opposite orientations of the
$\cal{PT}$-dipole, we calculate the difference in the intensity of scattered field between the configurations with $\cal{PT}$ dipoles aligned parallel and antiparallel to the direction of the incident wave.
\be
\label{deltaP}
\Delta P_{\rm S}(\vec{r}) = \left| | E_{\rm S}(\vec{r},\vec{p})|^2 - |E_S(\vec{r},-\vec{p})|^2 \right|.
\ee
We apply expressions (\ref{E_scatt_dipole},\ref{deltaP}) to the following two specific orientations of the dipole:

When the
$\cal{PT}$-dipole
 is oriented along the direction of the incident wave, $\vec{e_k}\parallel\Delta\vec{r}$, the scattering field can be expressed as a function of the observation angle $\theta$
\begin{equation}
\label{E_scatt_dipole_theta}
\begin{array}{lcl}
\displaystyle{E_{\rm S}(\vec{r})  = \frac{ 2ie^{i |k_0||\vec{r}|}}
{ |\vec{r}|^{{1}/{2}} }  }
\Big[ && \sr\cos(\frac{k_0a(1-\cos\theta)}{2})\\
&+&
\si\sin(\frac{k_0a(1-\cos\theta)}{2})\Big]
\end{array}
\end{equation}
where $\cos\theta = \vec{e}_k\cdot\vec{r}/|\vec{r}|$, and
\begin{equation}
\label{I_scatt_dipole_theta}
\Delta P_{\rm S}(\vec{r})  = \frac{ 4\sr\si}
{|\vec{r}|} \sin\left[k_0a(1-\cos\theta)\right].
\end{equation}
The behavior of the angular dependence  of $\Delta P_{S}(\vec{r})$, given by {\color{black}the} Eq. (\ref{I_scatt_dipole_theta}) is demonstrated in Fig. \ref{fig_1}.  The forward scattering does not depend on the orientation of the dipole ($\Delta P_{\rm S}(\theta=0)=0)$ while significant asymmetry is observed in the backward scattering ($\theta=\pi)$.
Interestingly, the backward scattering is symmetric in special cases when $2ak_0 = \pi\times N$, where $N$ is an integer, i.e. an "accidental" symmetric backscattering occurs at
the wavelengths $\lambda_N = 4a/N$.

In the case when the $\cal{PT}$-dipole is oriented {\color{black}perpendicularly} to the direction of the incident wave,
$\vec{e}_k\perp\Delta\vec{r}$, the scattering field can be expressed as a function of the observation angle $\theta$
\begin{equation}
\begin{array}{lcl}
\label{E_scatt_dipole_theta_perp}
\displaystyle{
E_{\rm S}(\vec{r})  = \frac{ 2ie^{i |k_0||\vec{r}|}}
{ |\vec{r}|^{{1}/{2}} }  }
\Big[ && \sr\cos(\frac{k_0a\sin\theta}{2}) \\
&-& \si\sin(\frac{k_0a\sin\theta}{2})\Big]
\end{array}
\end{equation}
and  the asymmetry in the scattered field between the opposite orientation of the $\cal{PT}$ dipole reads
\begin{equation}
\label{I_scatt_dipole_theta_perp}
\Delta P_{\rm S}(\vec{r})  = \frac{ 4\sr\si}
{|\vec{r}|} \sin\left(k_0a\sin\theta\right)
\end{equation}

\begin{figure}[t]
\noindent\includegraphics[width=0.4\textwidth]{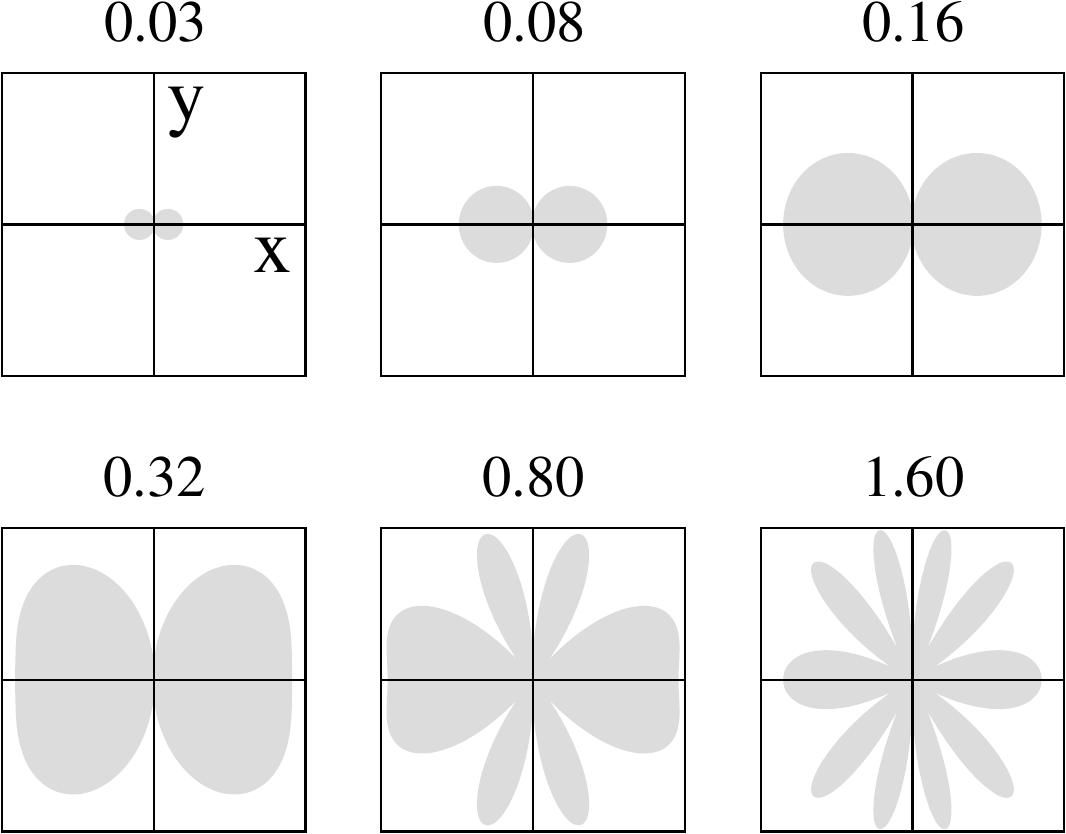}  
\caption{The same as in Fig. 2 but for
 perpendicular orientation of the $\cal{PT}$-dipole
}
\label{fig_1a}
\end{figure}

One can see in this case that no asymmetry between forward({\color{black}$\theta = 0$}) and backward({\color{black}$\theta = \pi$}) scattering occurs. In addition, the Eq. \ref{I_scatt_dipole_theta_perp} allows to determine the critical observation angle $\theta$ at which the asymmetry $\Delta P_{\rm S}(\vec{r})$ vanishes for a given frequency $f$: $\theta = \arcsin(N/(2f))$ and yields number of the critical observation angles $N_{\theta}$ which appear in one quadrant for the frequencies $f >  N_{\theta}/2$. These features arising from the Eq. (\ref{I_scatt_dipole_theta_perp}) can be observed in the angular dependence  of $\Delta P_{S}(\vec{r})$ shown in Fig. \ref{fig_1a}.

The  equations
(\ref{E_scatt_dipole},\ref{deltaP})
allow us to evaluate both frequency and angular dependence of the asymmetry scattering of the $\cal{PT}$ dipole and could be  be generalized for arbitrary orientation of the dipole.
As expected, scattered field decreases as $r^{-1}$ at large distances.
The only parameter  which determines the angular dependence of scattered field is $k_0a=2\pi a/\lambda$.

\section{Numerical method}
\label{numerics}

In numerical calculation,
the ${\cal{PT}}$-dipole is represented by two infinitely long, cylinders,
parallel to the $z$ axis.
The distance between centers of the cylinders is   $a$.
The  radius of cylinders is $R_0$ and refractive indices  $n_j = n_R \pm in_I$,  $j = 1,2$.
The incident electromagnetic plane wave with normalized frequency $f = a/\lambda$ propagating in the $xy$ plane
\be
E(x, y |\omega)_{\rm inc} =  \exp[i (k_x x+  k_y y) - i\omega t]
\label{eq.P1}
\ee
is polarized parallel to the axes of cylinders.

The total electric field can be expressed as the sum of the incident field $E(x,y|\omega)_{\rm inc}$ and a scattered field $E_{\rm S}(x,y|\omega)$
\be
E(x,y|\omega) =  E(x, y |\omega)_{\rm inc} +  E_{\rm S}(x,y|\omega)
\label{eq.P2}
\ee
To study scattering properties of EM waves for a single ${\cal{PT}}$-dipole we evaluate the radial component of the Poynting vector
\begin{equation}
P_{\rm S}(R,\phi) = E_{\rm S}(R,\phi)[H_{\rm S}^\phi(R,\phi)]^{*}
\label{eq.P3}
\end{equation}
along the circumference of the circle with radius $R$ centered at the focus of the system.
In {\color{black}the} Eq. (\ref{eq.P3}),
\be H^\phi_{\rm S} =  \frac{i}{\omega\mu }\frac{\partial E_{\rm S}}{\partial r}\Bigg|_{r=R}
\ee
is the tangential component of magnetic field.

In analogy to the Eq. (\ref {deltaP}) we characterize the asymmetry of scattered field associated with the opposite orientation of the
$\cal{PT}$-dipole in terms of the difference $\Delta P_{\rm S}(R,\phi)$ defined as
\be
\label{deltaPnum}
\Delta P_{\rm S}(R,\phi) = \left|  P_{\rm S}(R,\phi,\vec{p}) - P_S(R,\phi,-\vec{p}) \right|.
\ee

To compute the $\Delta P_{\rm S}(R,\phi)$  we apply a numerical  algorithm based on the expansion of electromagnetic field into cylinder functions \cite{Hulst}.
The scattered electric field  can be
expressed in cylindrical coordinates $r$ and $\phi$ as
\begin{equation}
E_{S}^{j}(\vec{r})  = \sum_{j=1}^{2}\sum_{m}\beta_m^{j}H_m(k_0|\vec{r}-\vec{r}_j|) e^{im\phi_j}.
\label{eq.1b}
\end{equation}
where $H_m$ are the Hankel functions of the first kind and $r_j$, $\phi_j$ are cylinder coordinates centered at the center of the  $j$th cylinder.
 The coefficients  $\beta_m$ can be calculated from the  continuity condition  of the tangential components of the electric and magnetic
 field at the boundary of cylinders.
Our approach is described in detail in Refs.~\onlinecite{PM3,PM4}.

\section{results}
\label{results}

\subsection{${\cal{PT}}$-dipole - parallel configuration}

\begin{figure}[t]
\noindent\includegraphics[width=0.23\textwidth]{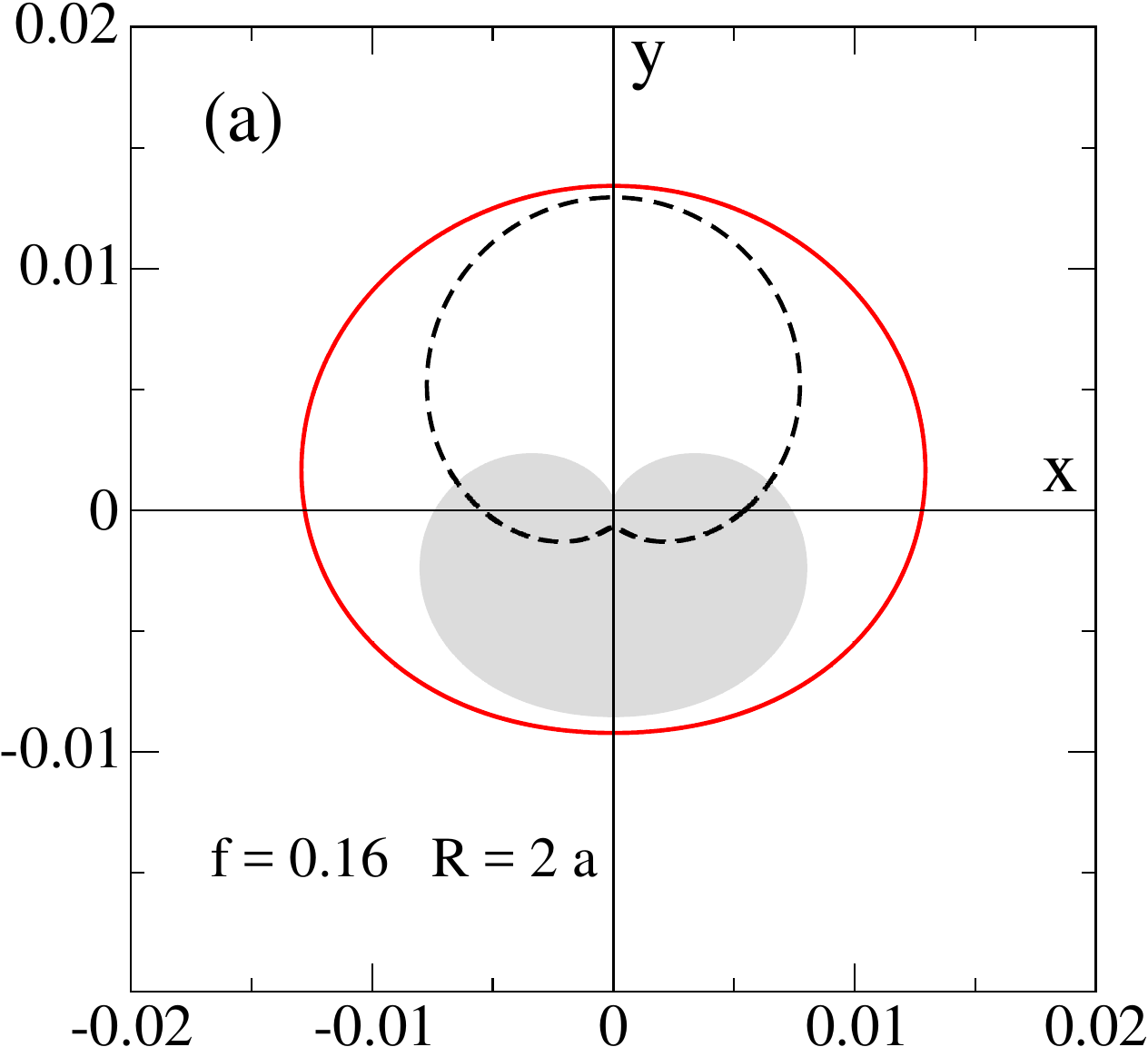}
~~
\noindent\includegraphics[width=0.23\textwidth]{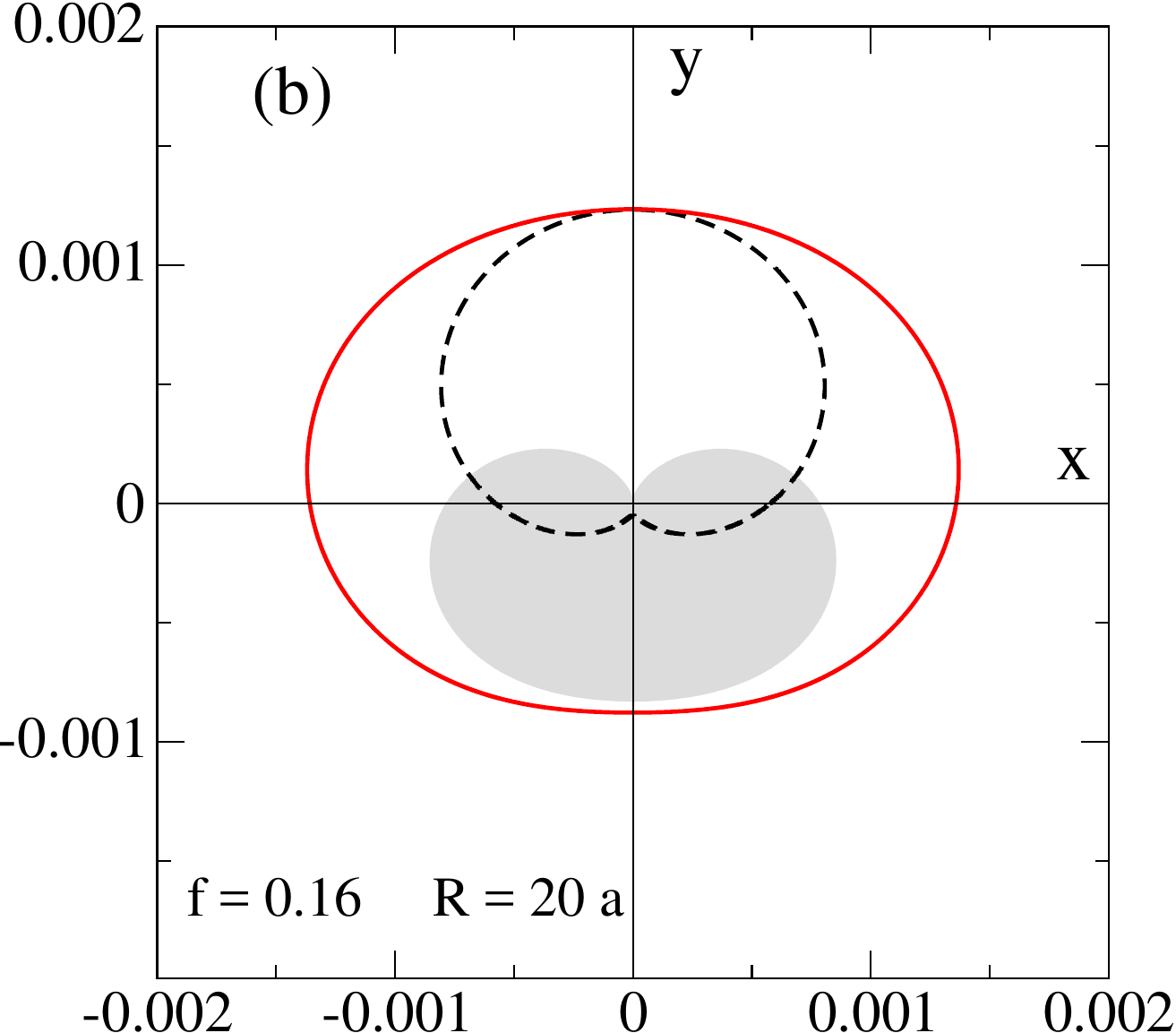}
\caption{(Color online) The intensity of scattered field for two opposite orientations of the dipole parallel to the propagation of the incident wave (dashed black and solid red lines)
and their difference  $\Delta P_{S}(R,\phi)$ (shaded area) for  the ${\cal{PT}}$-dipole with gain/loss characterized by $n_I = \pm 0.5i$.
(a) in the near field $(R = 2a)$, and (b) in the far field $(R = 20a)$. The frequency of an incident wave   $f = 0.16$.}
\label{fig_2}
\end{figure}

\begin{figure}[t]
\noindent\includegraphics[width=0.23\textwidth]{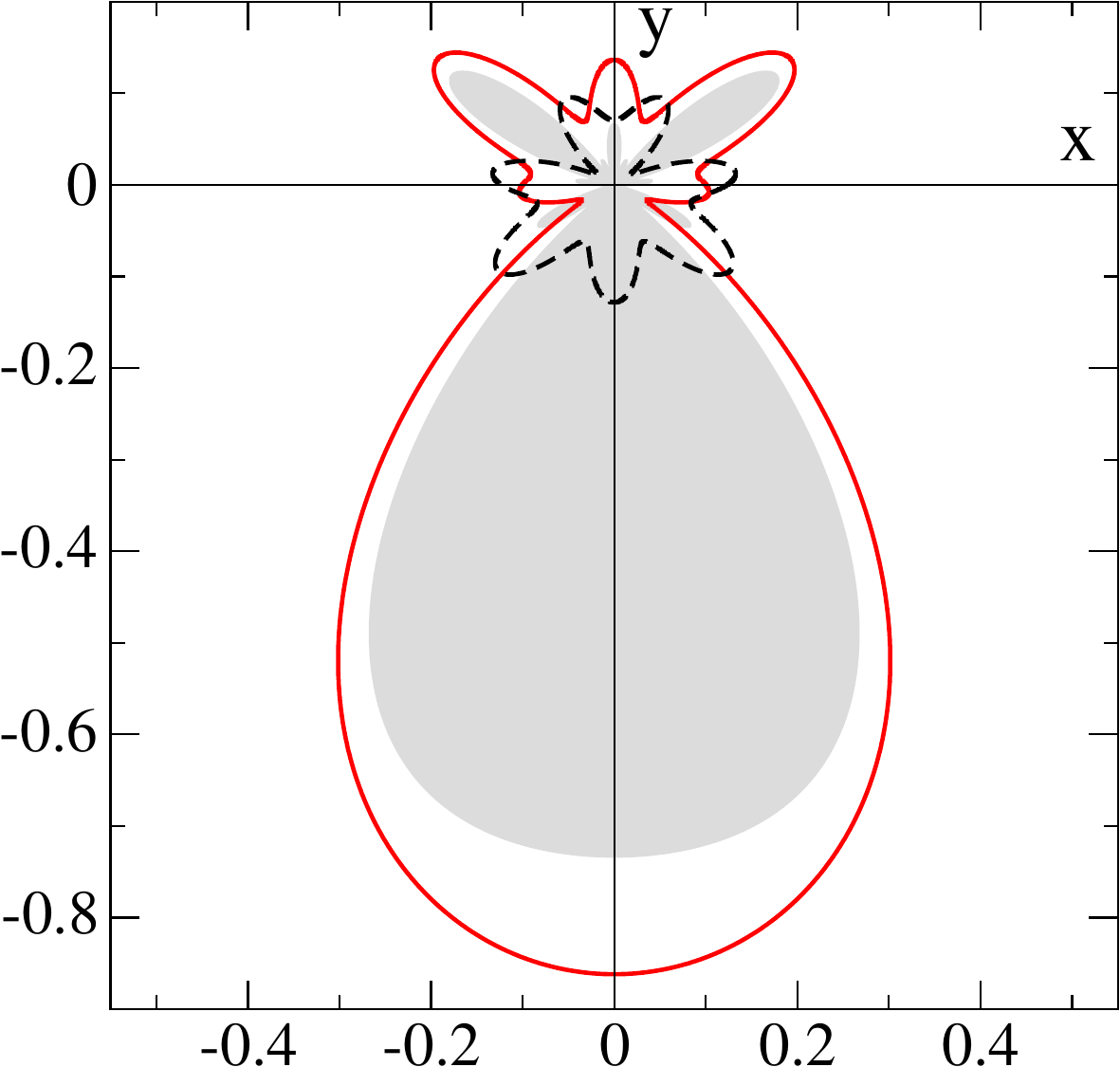}
~~
\noindent\includegraphics[width=0.23\textwidth]{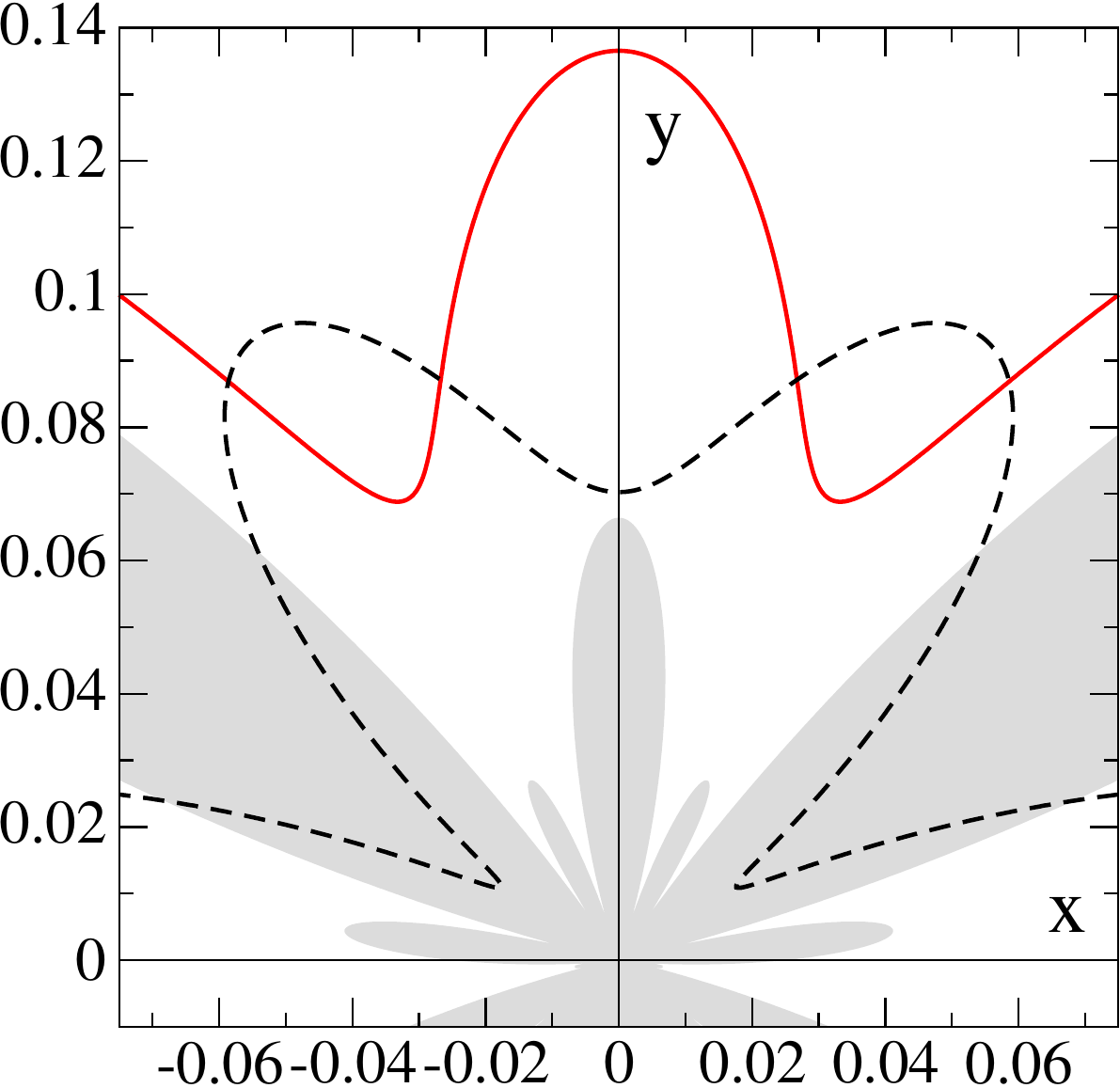}
\caption{(Color online) Left:  the difference in the intensity scattering for two antiparallel orientations of ${\cal{PT}}$-dipole $\Delta P_{S}(R,\phi)$ characterized by $n_I = \pm 0.5i$ in the near field $(R = 2a)$. Right panel shows the   detail of the scattered field.
 The frequency of an incident wave  $f = 1.0$.}
\label{fig_3}
\end{figure}

\begin{figure}[t]
\noindent\includegraphics[width=0.23\textwidth]{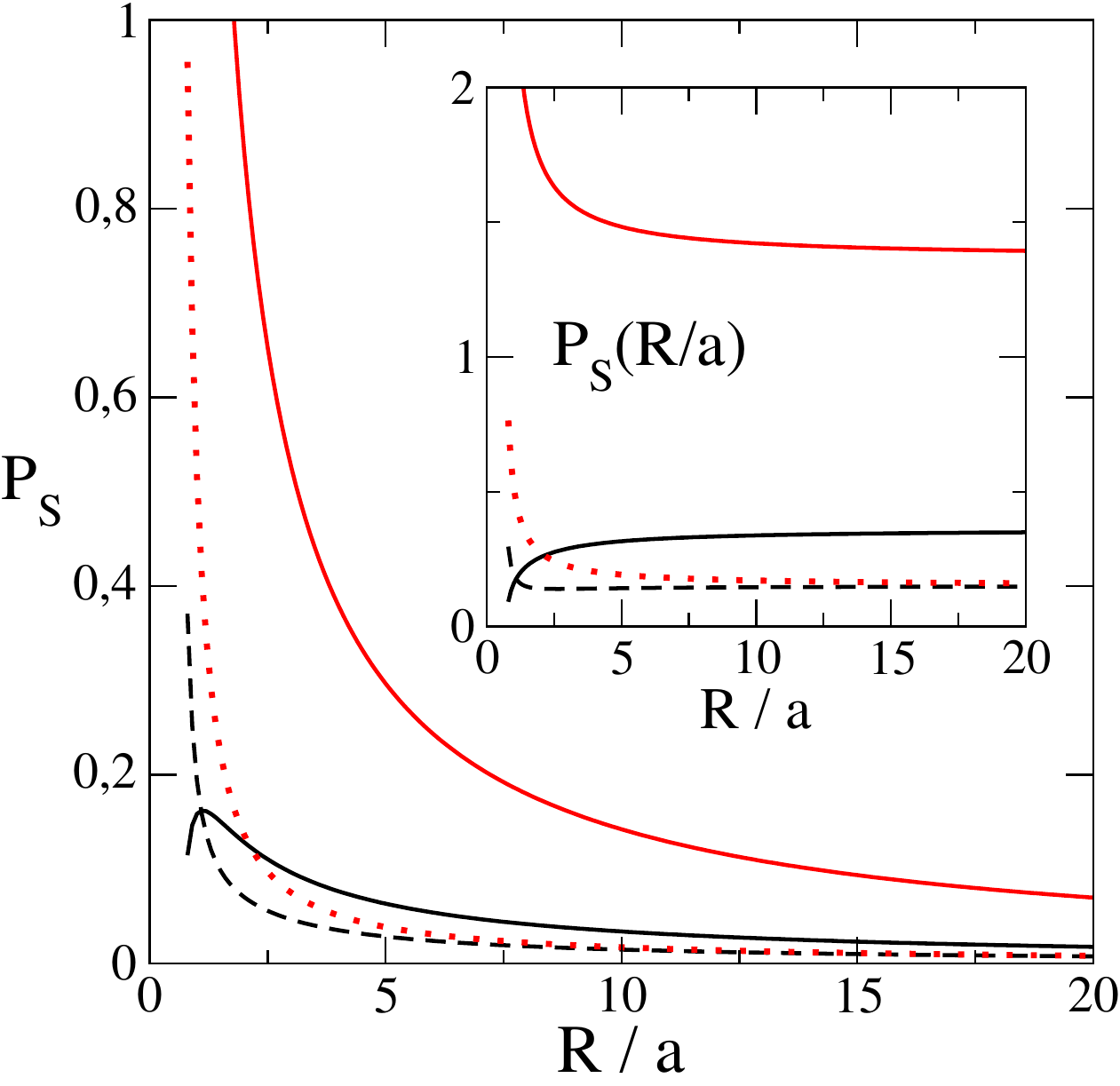}
%
\caption{(Color online) The intensity of the scattered field $P_{S}(R,\phi)$ along the $y$-axis as a function of the normalized radius $R/a$.
Solid lines represent backward scattering for two orientations of the dipole. Similarly,
black dashed and red dotted lines show  the forward scattering for two antiparallel orientations of the dipole.
Inset shows  the product $P_S(R/a)$ to prove that the intensity of scattered field decreases $\sim 1/R$ in the far field.
}
\label{fig_4}
\end{figure}

%

\begin{figure}[t]
\noindent\includegraphics[width=0.43\textwidth]{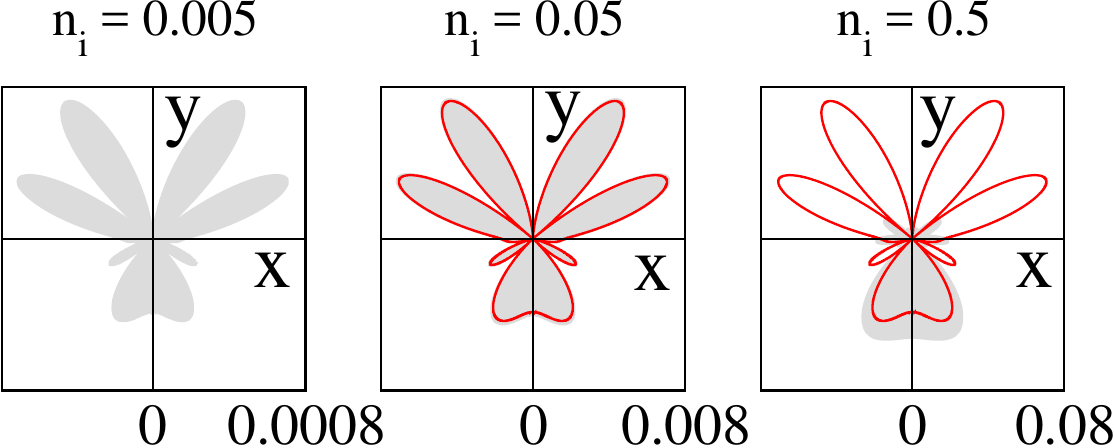}
%
\caption{(Color online) The difference in the intensity scattering for antiparallel orientations of ${\cal{PT}}$-dipole $\Delta P_{S}(R,\phi)$ in the far field($R = 20$) vs gain/loss parameter  $n_I = \pm 0.005i$ (left),  $n_I = \pm 0.05i$ (middle)  and  $n_I = \pm 0.5i$ (right),  when $f = 1.0$. Red line in the middle (right) panel  represents $\Delta P_S$
from the left (middle) panels, respectively, multiplied by factor of 10, to display a linear dependence  of scattered intensity
on the gain/loss parameter $n_I$ for small values $n_I$ and its breaking when $n_I$ increases.}
\label{fig_5}
\end{figure}

\begin{figure}[t]
\noindent\includegraphics[width=0.43\textwidth]{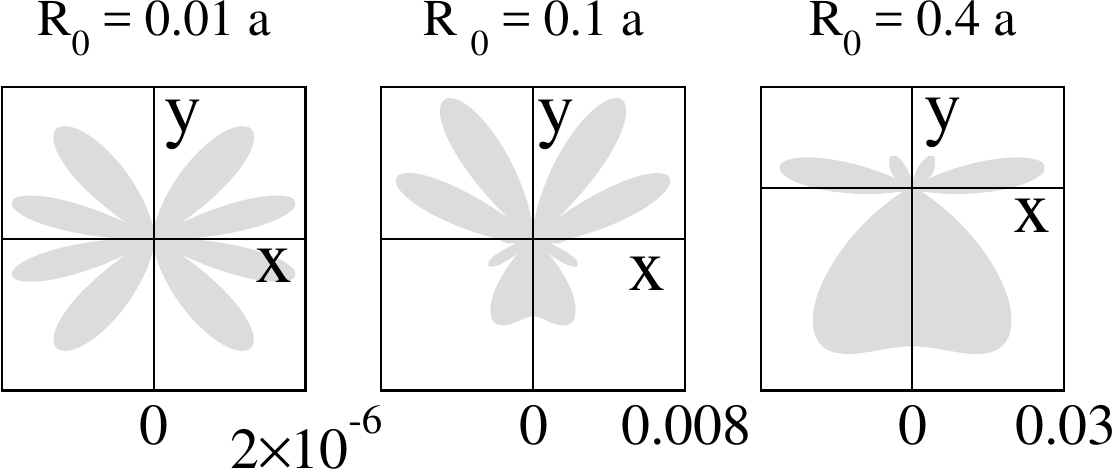}
%
\caption{(Color online) The difference in {\color{black}the intensity scattering} for antiparallel orientations of the ${\cal{PT}}$-dipole $\Delta P_{S}(R,\phi)$ in the far-field($R = 20$) vs radius of the cylinder for $n_I = \pm 0.05i$ $R_0 = 0.01a$ (left), $R_0 = 0.1a$ (middle), and  $R_0 = 0.4a$ (right). The frequency $f = 1.0$. Note that Born approximation predicts the symmetrical backscattering for this frequency. Numerically, this is observed   only for very tiny cylinders (the left panel). For stronger scatterers, Born approximation is not valid.}
\label{fig_6}
\end{figure}

We first consider the case when the ${\cal{PT}}$-dipole is parallel to the propagation direction of the  incident wave. The system which represents a $\cal{PT}$-dipole consists of two cylinders of radius $R_0 = 0.1a$ characterized by
the refractive index  $n_i = n_R \pm in_I$  $i = 1,2$, where real part $n_R = 3.5$ is kept constant while an imaginary part is varied in the range $0.005 < n_I < 0.5$.
We have chosen the radius of the cylinder to be sufficiently small $R_0 = 0.1a$ to allow comparison with  analytic results based on the point scatterer approximation.
In Fig. \ref{fig_2}. we display the scattering diagrams obtained for the frequency $f = 0.16$ for two
the parallel orientations of the dipole $\pm \vec{p}$  in the  near-field ($R=2a$) and far-field limit($R = 20a$).
The grey shaded area in Fig. \ref{fig_2} which shows the absolute value of the difference between the scattered power $\Delta P_{S}(R,\phi)$ for both orientations $\vec{p}$ and $-\vec{p}$ represent the key feature associated with scattering properties of the $\cal{PT}$ dipole.
Primarily, in the case of parallel orientation of the $\cal{PT}$ dipole the scattering diagrams reveal a strong asymmetry along the direction of propagation of the incident wave. Such a behavior is consistent with analytical results given by the Eq. \ref{I_scatt_dipole_theta} and it is displayed in Fig. \ref{fig_1}.
In addition, one can observe that the in the $\it{far}$-$\it{field}$ the power scattered along the $y-$axis for two antiparallel orientations of the dipole shown in Fig. \ref{fig_2}(b) coincide and yield a vanishing difference $\Delta P_{S}(R,\phi=0)$.  Simultaneously, this behavior confirms the theoretical prediction given by the {\color{black}Eq.
\ref{E_scatt_small-dipole}}.
By inspecting the scattering of the ${\cal{PT}}$-dipole in the ${\it near}$-${\it field}$  limit we found the
asymmetry   of the transmitted power along $y-$axis as it is demonstrated in Fig. \ref{fig_2}(a).
We explored the existence of this phenomenon also at larger frequencies where the system reveals according to theoretical model richer scattering patterns (Fig. \ref{fig_1}). As an example  we display
in Fig.~\ref{fig_3}
the scattering diagram for the frequency $f=1$.

To quantify the transition between the near and far field limit we depict in Fig. \ref{fig_4} the dependence of the field scattered along the $y-$axis on the normalized radius $R/a$ for both orientations of the dipole. One can see that the asymmetry  in the forward scattering vanishes at $R \simeq 12.5a$ which {\color{black}suggests} that the non-vanishing difference in the forward scattering appears solely in the near-field regime. Since scattered fields decrease as $\sim 1/R$ in the far field, the normalized product $RP/a$ remains constant for the backward scattering as it is shown in the inset of Fig. \ref{fig_4}.

Besides the effect associated with near-field asymmetry described above, we found yet another interesting difference between the theoretical prediction and numerical results.
Namely, we observed that the scattering pattern associated with {\color{black}a  ${\cal{PT}}$-dipole} in the far-field limit obtained numerically reveals strong dependence on the strength of the imaginary part $n_I$. It is demonstrated in Fig. \ref{fig_5}, where dependencies of the scattering diagrams for three values of the gain/loss parameters are depicted. We note that according to the theoretical model the difference between the the intensities for antiparallel orientations of the ${\cal{PT}}$-dipole given by {\color{black}the} Eq.
\ref{I_scatt_dipole_theta}, the size of the imaginary component $n_I$ {\color{black}does not affect} the shape of the $\Delta P_{S}(R,\phi)$. The theoretical approach in principle cannot account for the features described above.

To quantify the range of the gain/loss parameter beyond which the numerical results indicate limits of the validity of the Born approximation we compare  in  the middle  panel of Fig. \ref{fig_5}  the scattering intensities for
two values of $n_I$: 0.005 (red line, multiplied by 10)  and $n_I=0.05$. Clearly, the scattering increase linearly with $n_I$ for small values of $n_I$.
The same procedure  applied to the results for $n_I=0.05$ and 0.5 (right panel of Fig. \ref{fig_5})
{\color{black}unveils} the breaking of linear behavior for higher gain/loss parameter.

For completeness, we also studied how the scattering pattern is affected when the radius of the cylinder is varied in the range $0.01a < R_0 < 0.4a$ -- see   Fig. \ref{fig_6}. One can observe that in comparison with the results shown in Fig. \ref{fig_5} which display the dependence of the $\Delta P_{S}(R,\phi)$ on the $n_I$,  the variation of the radius gives rise to a significantly wider range of the $\Delta P_{S}(R,\phi)$ and to a strong dependence of the shape of the scattering pattern.  We do not expect any simple $R_0$ dependence of the scattering pattern since the latter is strongly affected by Fano resonances \cite{rr,PM4}.

The result shown in left panel of Fig. \ref{fig_6} confirms the accidental symmetry in the {\color{black} backscattering} in the far-field limit along the $y$-axis which occurs
	for small $\cal{PT}$ dipoles. The effect observed in numerical calculations occurs at the integer-valued frequencies at which the difference $\Delta P_{\rm S}(R,\phi)$ vanishes and {\color{black}is in accord with} the theoretical model in the far-field limit given by {\color{black}the} Eq. \ref{I_scatt_dipole_theta}. This result also indicates that Born approximation {\color{black} is} not sufficient for thicker cylinders as it shown in {\color{black}the} middle and {\color{black}the} right panels of {\color {black} Fig.} \ref{fig_6}, where the backscattering is not symmetric.

\subsection{${\cal{PT}}$-dipole - perpendicular configuration}

\begin{figure}[t]
\noindent\includegraphics[width=0.23\textwidth]{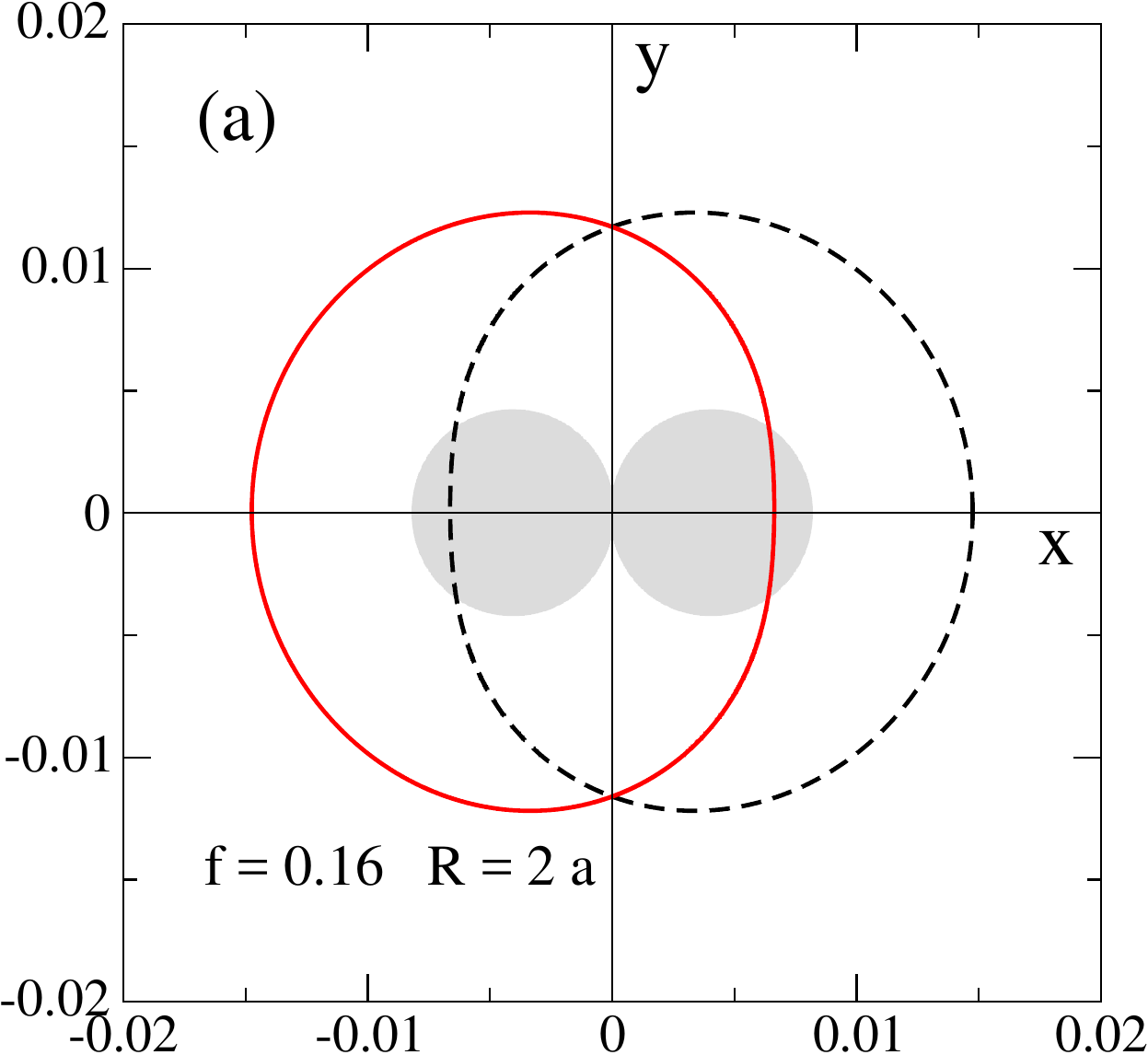}
~~
\noindent\includegraphics[width=0.23\textwidth]{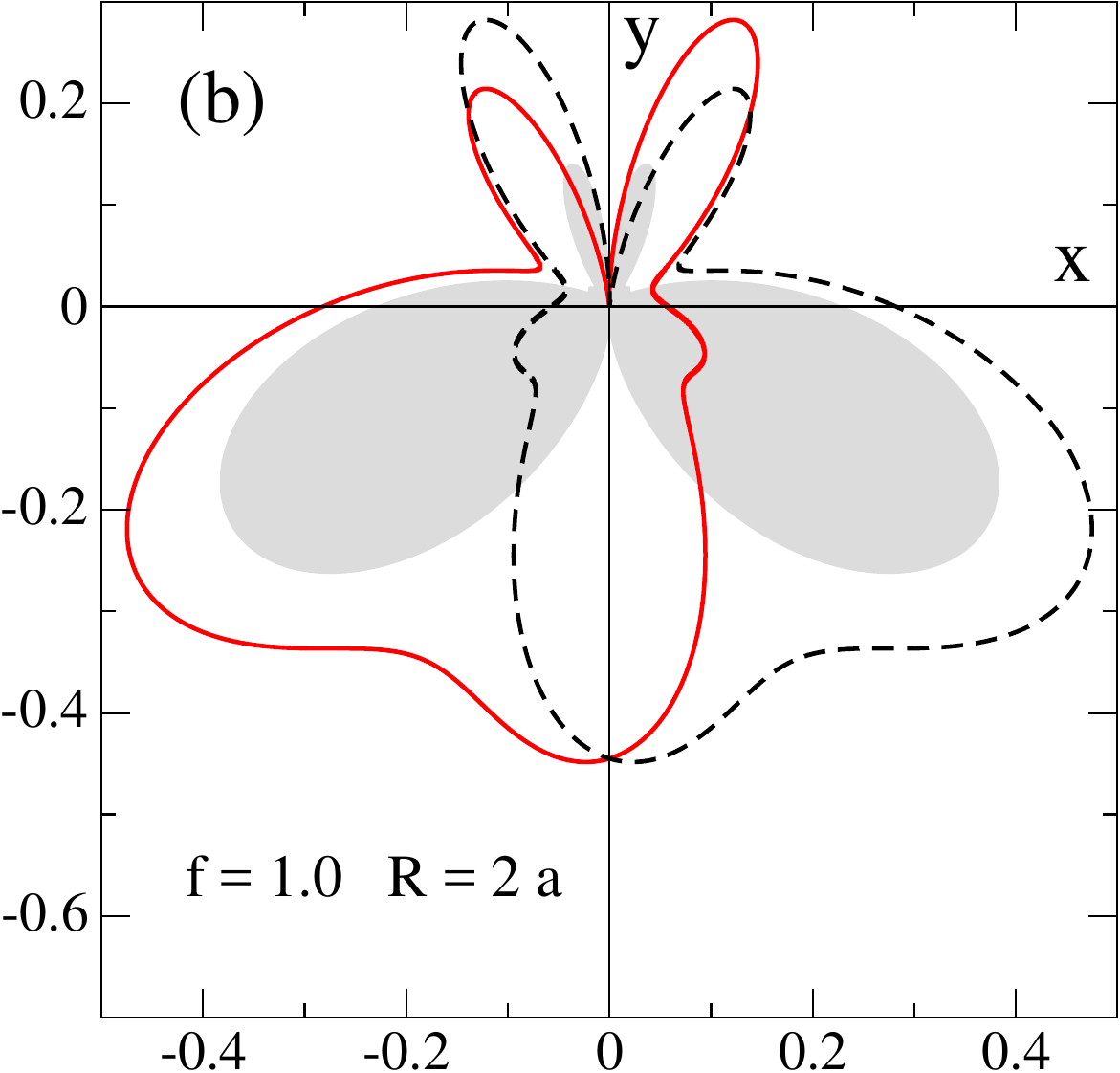}
\caption{The difference in the intensity scattering for two antiparallel orientations of ${\cal{PT}}$-dipole lying perpendicularly to incident wave.
(black dashed and red solid line).
Dashed area is the difference $\Delta P_S$.
Gain/loss parameter  $n_I = \pm 0.5$. Left:
$f = 0.16$ , right:  $f = 1.0$.
}
\label{fig_7}
\end{figure}

\begin{figure}[t]
\noindent\includegraphics[width=0.43\textwidth]{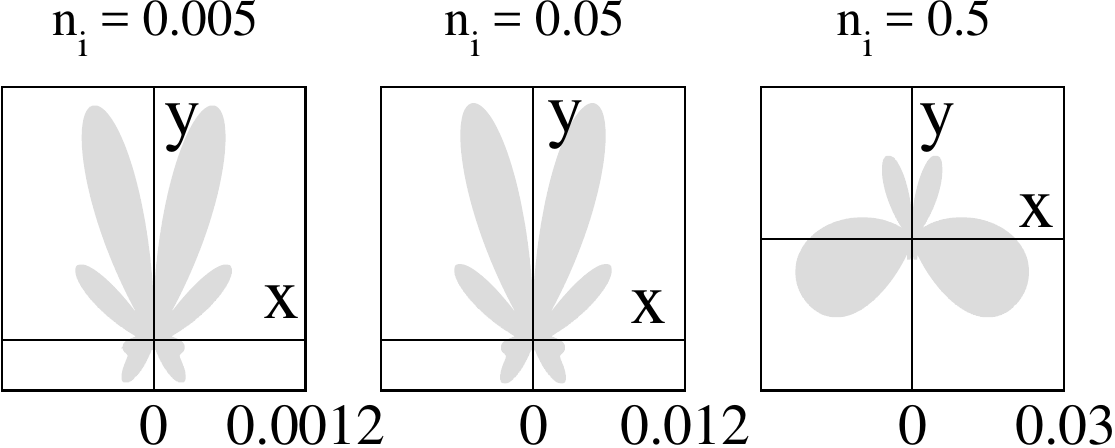}
%
\caption{Dependence of the difference in the intensity scattering for antiparallel orientations of ${\cal{PT}}$-dipole given by $\Delta P_{S}(R,\phi)$ in the far field($R = 20$) on the gain/loss parameter for  $n_I = \pm 0.005i$ (left), $n_I = \pm 0.05i$ (middle)  and  $n_I = \pm 0.5i$ (right),  when $f = 1.0$.}
\label{fig_8}
\end{figure}

In Fig. \ref{fig_7} we show the scattering patterns in the near-field limit($R = 2a$) for the perpendicular orientation of the ${\cal{PT}}$-dipole for two different frequencies. The intensity of the scattered power $\Delta P_{S}(R,\phi)$ for parallel and antiparallel orientations $\pm\vec{p}$ indicated by dashed black and solid red lines possess the features which significantly deviate from those associated with parallel orientation of the $\cal{PT}$-dipole while the asymmetric scattering indicated by the shaded areas is maintained. It is interesting to note that for small frequencies the $\Delta P_{S}(R,\phi)$  is symmetric along both $x$ and $y$-axis  while with an increasing frequency becomes strongly asymmetric along the $y$-axis. One can also observe that in contrast to the parallel orientation of the ${\cal{PT}}$-dipole asymmetric scattering does not occur along the $y$-axis in accord with theoretical model -- see {\color {black} Eq. \ref{I_scatt_dipole_theta_perp}.}

In Fig. \ref{fig_8} we demonstrate the dependence of the scattering diagrams in the far-field for three values of the gain/loss parameter $n_I$. The scattering patterns display qualitatively
similar behavior as those associated with parallel orientation shown in Fig. \ref{fig_6}, in particular they confirm a linear dependence on the gain/loss parameter $n_I$ in the range
$0.005 < n_I < 0.05$ within the range of the validity of the first Born approximation.

\section{Discussion and Conclusions}

\begin{figure}[t]
\noindent\includegraphics[width=0.4\textwidth]{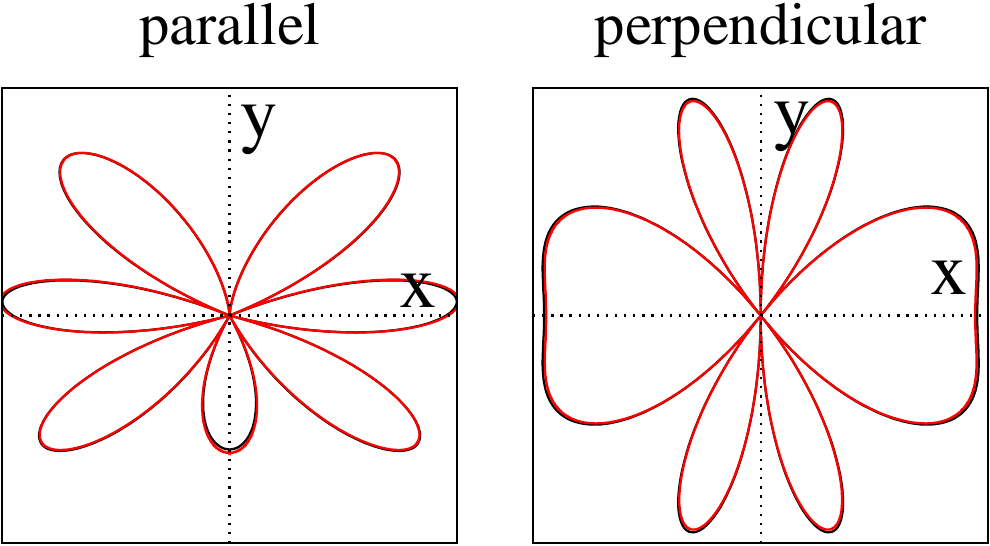}
\caption{(Color online) Angular diagrams of $\cal{PT}$ asymmetry for both parallel and perpendicular orientation of the $\cal{PT}$-dipole given by $\Delta P_{S}(R,\phi)$ at frequency $f=0.8$. Analytical prediction, given by Eq. \ref{I_scatt_dipole_theta} (black line)
is compared with numerical data for $R_0=0.01a$, $n_I=0.005$ and $R=20a$ (red line, re-scaled in absolute value).
}
\label{KS-fig5}
\end{figure}

The numerical results presented in the previous Section confirm the  asymmetric scattering of the  ${\cal{PT}}$-dipole for both parallel an perpendicular dipole orientations  and simultaneously offer possibility to examine the limits of the theoretical model associated with the approximations implemented.

First of all, we numerically explored differences in the scattering patterns arising in the near-field which are demonstrated in Figs. \ref{fig_3} and \ref{fig_4}, in particular that in the near-field the scattering {\color{black}in} the forward direction of ceases to be symmetric. This feature clearly arises due to the modified properties of the scattering pattern when the transition between near and far-field limit takes place as it has been shown in Fig. \ref{fig_4} and yields the threshold between both regimes.

To check the validity of Born approximation we compare the results obtained analytically  and numerically for the case of small $\cal{PT}$-dipole
with the radius $R = 0.01a$ and gain/loss parameter $n_I = 0.005$ -- see Fig. \ref{KS-fig5}. The analytical and numerical results coincide for both orientation of the $\cal{PT}$-dipole.

In addition, the limit of the validity of the first Born approximation {\color{black}has} been determined by exploring the dependence of the scattering of the  ${\cal{PT}}$-dipole on the size of the gain/lass parameter $n_I$. We have shown that when the size of the imaginary component $n_I$ is small ($\lesssim 0.05$) {\color{black}it} does not qualitatively affects the shape of the $\Delta P_{\rm S}(R,\phi)$ and follows the linear dependence on the $n_I$ in accord with the first Born approximation. When the size of the $n_I$ is increased,  the linear scaling of the scattering with the $n_I$ does not apply and the system cannot be described in terms of the first Born approximation.
Finally, we note that the results shown in Fig. \ref{fig_7} which demonstrate the dependence of the $\Delta P_{S}(R,\phi)${\color{black},} display the strong dependence on the radius of the cylinder $R_0$, however one cannot anticipate any trivial scaling since its behavior may be strongly affected for the frequencies in the vicinity of the Fano resonances.

In conclusion, we analyzed, both analytically and numerically, the electromagnetic response of the ${\cal{PT}}$ dipole
and found that the Born approximation is valid  in the limit of far-field and tiny scattering parameters of the dipole. For a general case,
rich variety of the scattering pattern is observed both for the parallel and perpendicular orientation of the dipole.
Our results indicate that structures composed from large number of ${\cal{PT}}$ dipoles might possess  interesting new transmission properties, worth to be analyzed in the future.

\section*{Acknowledgements}
We acknowledge financial support by Spanish Ministerio de Ciencia e Innovaci{\'o}n, the European Union FEDER through project FIS2011-29731-C02-01. The research of P.~Marko\v s was supported  by the Slovak Research and Development Agency under the contract No. APVV-15-0496
and by the Agency  VEGA under the contract No. 1/0108/17. The research of V. Kuzmiak was supported by
Grant 16-00329S of the Czech Science Foundation(CSF).

\end{document}